\documentclass[
aps,
prb,
showpacs,
floatfix,
reprint,
twoside,
superscriptaddress
]{revtex4-1}
\usepackage{amsmath}
\usepackage[T1]{fontenc}
\usepackage{fourier}
\usepackage{graphicx}
\usepackage[colorlinks=true,
allcolors=blue
]
{hyperref}

\begin{document}

\title{Hidden Order as a Source of Interface Superconductivity}

\author{Andreas Moor}
\affiliation{Theoretische Physik III, Ruhr-Universit\"{a}t Bochum, D-44780 Bochum, Germany}
\author{Anatoly F.~Volkov}
\affiliation{Theoretische Physik III, Ruhr-Universit\"{a}t Bochum, D-44780 Bochum, Germany}
\author{Konstantin B.~Efetov}
\affiliation{Theoretische Physik III, Ruhr-Universit\"{a}t Bochum, D-44780 Bochum, Germany}
\affiliation{National University of Science and Technology ``MISiS'', Moscow, 119049, Russia}

\begin{abstract}
Interfacial superconductivity is observed in a variety of heterostructures composed of different materials including superconducting and nonsuperconducting (at appropriate doping and temperatures) cuprates and iron-based pnictides. The origin of this superconductivity remains in many cases unclear. Here, we propose a general mechanism of interfacial superconductivity for systems with competing order parameters. We assume that parameters characterizing the material allow formation of another order like charge- or spin-density wave competing and prevailing superconductivity in the bulk (hidden superconductivity). Diffusive electron scattering on the interface results in a suppression of this order and releasing the superconductivity. Our theory is based on the use of Ginzburg--Landau equations applicable to a broad class of systems. We demonstrate that the local superconductivity appears in the vicinity of the interface and the spatial dependence of the superconducting order parameter~$\Delta(x)$ is described by the Gross--Pitaevskii equation. Solving this equation we obtain quantized values of temperature and doping levels at which~$\Delta(x)$ appears. Remarkably, the local superconductivity shows up even in the case when the rival order is only slightly suppressed and may arise also on the surface of the sample (surface superconductivity).
\end{abstract}

\date{\today}

\maketitle

\section{Introduction}

Interesting phenomena have been discovered few years ago in the study of superconductivity in different materials, especially in high\nobreakdash-$T_{\text{c}}$ superconductors---cuprates and Fe\nobreakdash-based pnictides. It turned out that the critical temperature of the superconducting transition~$T_{\text{c}}$ in heterostructures, e.g., in bilayers, is higher than the critical temperature~$T_{\text{c}}$ of bare films that can be even nonsuperconducting.\cite{Reyren_et_al_2007,Gozar_et_al_2008,Caviglia_et_al_2008,Bell_et_al_2009,Kozuka_et_al_2009,Biscaras_et_al_2010,Logvenov_Gozar_Bozovic_2009,Smadici_et_al_2009,Koren_Milo_2010,Loktev_Pogorelov_2008,Butko_et_al_2009,Bergman_Pereg_Barnea_2011,Pereiro_et_al_2011} The authors of Ref.~\onlinecite{Yuli_et_al_2008} (see also Ref.~\onlinecite{Koren_Milo_2010}) studied bilayers consisting of two cuprates---overdoped (La$_{2-x}$Sr$_{x}$CuO$_{4}$ with ${x = 0.35}$) and underdoped (${0.1 < x < 0.12}$) cuprate films. The largest increase of~$T_{\text{c}}$ was about~$11$~K---from ${T_{\text{c}} = 21~\text{K}}$ in bare underdoped films to ${T_{\text{c}} = 32~\text{K}}$ in bilayers. Moreover, superconductivity has been observed in bilayers composed of nonsuperconducting materials, La$_{2-x}$Sr$_{x}$CuO$_{4}$ with ${x = 0}$ (an insulator) and La$_{2-x}$Sr$_{x}$CuO$_{4}$ with ${x = 0.45}$ (normal metal).\cite{Gozar_et_al_2008} The critical temperature reached~$50~\text{K}$. Further experimental studies showed the independence of the critical transition temperature in bilayers La$_{2-x}$Sr$_{x}$CuO$_{4}$/La$_{2}$SrCuO$_{4}$ on~$x$ in a rather wide doping interval (${0.15 < x < 0.47}$).\cite{Wu_et_al_2013}

A slight increase of~$T_{\text{c}}$ has been measured also in another high\nobreakdash-$T_{\text{c}}$ superconductor---YBa$_{2}$Cu$_{3}$O$_{7}$~films covered by a thin Ag film.\cite{Tinchev_2010} A few decades ago, a similar effect has been observed in bilayers composed of conventional low\nobreakdash-$T_{\text{c}}$ superconductors and normal metals.\cite{Ruhl_1965,Naugle_1967,Strongin_et_al_1968} These latter experiments were motivated by original Ginzburg's ideas on the possibility to get a surface superconductivity with a high transition temperature.\cite{Ginzburg_1964,Allender_et_al_1973}

Another high\nobreakdash-$T_{\text{c}}$ superconductors where an enhancement of superconductivity at the interface has been established are the so-called iron-based pnictides discovered in~2008.\cite{Hosono_2008} In this type of superconductors competing order parameters~(OP), magnetic and superconducting, may coexist. To be more exact, the spin density wave~(SDW) and the superconducting~OP may arise in these materials (see reviews Refs.~\onlinecite{Stewart_2011,Wen_Li_2011,Hirschfeld_Korshunov_Mazin_2011,Norman_2008,Chubukov_2012}) with the amplitudes~$W$ and~$\Delta $ depending on temperature and doping. The interfacial superconductivity in one of the Fe\nobreakdash-based pnictides (CaFeAs doped with La, Ce, Pr or Nd) was observed by Wei~\emph{et al.}\cite{Wei_et_al_2013} Whereas the bulk critical temperature was equal to about~$30~\text{K}$, a small fraction of samples had a ${T_{\text{c}} \simeq 49~\text{K}}$. Even higher ${T_{\text{c}} \simeq 77~\text{K}}$ was achieved by a Chinese experimental group in another Fe\nobreakdash-based material (single unit-cell FeSe films on SrTiO$_{3}$).\cite{Wang_et_al_2012}

Very encouraging for understanding the very nature of high-$T_{\text{c}}$ superconductivity seems to be the effect of apparent enhancement of superconducting transition temperature at the interface between an iron-based chalcogenide superconductor (FeSe) and SrTiO$_{3}$ used as substrate.\cite{Xiang_et_al_2012,Tan_et_al_2013} This discovery questioned the role of phonons in bulk iron-based superconductors~\cite{Xiang_et_al_2012} and confirmed the presence of the magnetic order (spin-density wave) as a key ingredient for high-$T_{\text{c}}$ superconductivity in iron-based superconductors.\cite{Tan_et_al_2013}

Rather actively the interface superconductivity is studied in heterostructures LaAlO$_{3}$/SrTiO$_{3}$.\cite{Reyren_et_al_2007} It is assumed
that a two-dimensional electron gas is formed at the interface. Effect of an electric field has been employed to explore the phase diagram of LaAlO$_{3}$/SrTiO$_{3}$ interface.\cite{Caviglia_et_al_2008,Bell_et_al_2009} Aside from superconductivity, the phenomenon of ferromagnetism induced at the interface of an oxide heterostructure has been observed recently.\cite{Salluzzo_et_al_2013,Li_et_al_2011} The review in Ref.~\onlinecite{Hwang_et_al_2012} provides an excellent overview over the possible symmetries and degrees of freedom of correlated electrons that can evolve at oxide interfaces. It includes, inter alia, superconductivity, magnetism, ferroelectricity, and charge- and spin orders as well. Moreover, it has been found, that at the interfaces between LaAlO$_{3}$ and SrTiO$_{3}$, superconductivity coexists with ferromagnetism,\cite{Dikin_et_al_2011,Julie_et_al_2011,Li_et_al_2011,Fidkowski_et_al_2013} a surprising result offering a potential for exotic superconducting phenomena due to highly broken inversion symmetry of the interface and a ferromagnetic background.\cite{Julie_et_al_2011}

Several theories have been suggested to explain the phenomenon of interface superconductivity. Some of them consider a nonuniform charge distribution near the interface and use a phenomenological relation of~$T_{\text{c}}$ to this distribution.\cite{Logvenov_Gozar_Bozovic_2009,Loktev_Pogorelov_2008,Loktev_Pogorelov_2010} Different ideas have been used in other theories,\cite{Berg_et_al_2008} where bilayers composed of superconductors with different ratio of the~$T_{\text{c}}$ and pairing strength were considered and it was assumed that~$T_{\text{c}}$ is suppressed by phase fluctuations. In the vicinity of the interface the role of these fluctuations is not important as compared to the bulk due to suppression of fluctuations by the proximity effect. However, this suggestion cannot explain the observed independence of~$T_{\text{c}}$ on the doping level~$x$. Perhaps, there is no single mechanism responsible for the interface superconductivity because it was observed in quite different materials under various conditions. The important ingredient of this effect is the presence of an interface or some sort of nonhomogeneity.

In this paper we propose a new mechanism for the interface superconductivity. The proposed mechanism is very robust and general being independent of the microscopic details of considered materials. It is applicable to any materials where, alongside with the superconducting order parameter~(OP), another OP exists. We do not pretend to apply our theory to any system where the interface superconductivity occurs, but we show that it can be used to materials in which two OPs may potentially exist. As is well known, in high\nobreakdash-$T_{\text{c}}$ superconductors, an important role is played by the charge- or spin-ordering.\cite{Stewart_2011,Wen_Li_2011,Hirschfeld_Korshunov_Mazin_2011,Norman_2008,Chubukov_2012} In cuprates, a charge density wave~(CDW) has been observed
recently in numerous experiments\cite{Lawler_et_al_2010,Parker_et_al_2010,Julien_2011,Julien_2013,Ghiringhelli_et_al_2012,Chang_et_al_2012,Achkar_et_al_2012,LeBoeuf_et_al_2013,Blackburn_et_al_2013,Comin_et_al_2014,Fujita_et_al_2014} and discussed in Refs.~\onlinecite{Vojta_Rosch_2008,Metlitski_Sachdev_2010,Sachdev_La_Placa_2013,Meier_et_al_2014,Wang_Chubukov_2014}. It may exist alongside with superconductivity, whereas in Fe\nobreakdash-based superconductors, the spin density wave is more important. The presence of the nonsuperconducting OP~$W$ (CDW or SDW) changes such characteristics of superconducting state as London penetration depth,\cite{ExperLondon,ChubukovLondon,SachdevLondon} heat capacity,\cite{ExperHeat,Levchenko14} etc.

We show that the presence of a `hidden' OP~$\Delta$ in a heterostructure with ${W \neq 0}$ may lead to appearance of local superconductivity at the interface at temperatures ${T > T_{\text{c}}(W)}$, where~$T_{\text{c}}(W)$ is the critical temperature of the superconducting transition of bare films composing the heterostructure which depends on the amplitude~$W$. We consider a heterostructure with an interface where~$W$ is locally suppressed. This suppression may be caused by an enhanced impurity scattering and doping level in the vicinity of the interface. The enhanced impurity scattering can be caused by the interdiffusion of atoms and/or roughness of the interface. Both factors suppress the CDW or SDW. In this case, in a vicinity of the interface where~$W$ is suppressed, local superconductivity arises with~$\Delta(x)$ decaying on a characteristic scale of the order of the superconducting coherence length~$\xi_{\text{s}}$. Interestingly, the local superconductivity occurs at `quantized' temperatures because the OP~$\Delta(x)$ is described by the linearized Gross--Pitaevskii equation, i.e., by the Schr\"{o}dinger equation with a one-dimensional potential well which always has discrete energy levels~(or only one level). In one-dimensional case (flat interface) the local superconductivity arises even at a rather small suppression of~$W$.

Note that stimulation of the bulk superconductivity by impurities in materials with two OPs was considered by one of the authors a long time ago.\cite{Efetov_1981} Recently, the effect of superconductivity stimulation by impurities in the bulk of Fe\nobreakdash-based pnictides with two OPs has been analysed by Fernnades~\emph{et~al.} in Ref.~\onlinecite{Fernandes_Vavilov_Chubukov_2012}. However, to the best of our knowledge, there has been no study of the mechanism of the interface superconductivity presented in this paper.

In Section~\ref{sec:General}, the general nonhomogeneous Ginzburg--Landau equations describing two coupled order parameters are introduced. In the homogeneous case, we derive conditions for the coefficients in the Ginzburg--Landau equations that should be fulfilled for obtaining one of the possible three phases: 1)~pure superconducting, 2)~pure charge- or spin-density wave, 3)~a mixed state. In the nonhomogeneous case, a solution of the Ginzburg--Landau equations for the order parameters is obtained, whereby detailed calculations are shifted to the Appendix. In Section~\ref{sec:Aplication}, we discuss applicability of the theory to the case when the constituents of the heterostructure are high\nobreakdash-$T_{\text{c}}$ cuprates or iron-based pnictides. We propose also an experimental setup suitable for testing our predictions. Concluding, we discuss our results in Section~\ref{sec:Discussion}.

\section{General considerations}
\label{sec:General}

\subsection{Free energy and self-consistency equations}

We start with the expression for the free energy~$\mathcal{F}$ for a system with two order parameters~(OPs),~$\Delta$ and~$W$. In one-dimensional case, i.e., in case of a preferred direction provided by the interface, the free energy is given by
\begin{align}
\mathcal{F} = \frac{1}{2} \int \mathrm{d} x & \Bigg[ \xi_{\text{s}}^{2} (\Delta^{\prime})^{2} - a_{\text{s}} \Delta^{2} + \frac{b_{\text{s}}}{2} \Delta^{4} + \gamma \Delta^{2} W^{2} + \label{eq:free_energy} \\
& + \xi_{\text{w}}^{2} (W^{\prime})^{2} - a_{\text{w}} W^{2} + \frac{b_{\text{w}}}{2} W^{4} \Bigg] \,, \notag
\end{align}
where~$\Delta^{\prime}$ and~$W^{\prime}$ denote the spatial derivatives of corresponding order parameters, and the coefficients~$\xi_{\text{s},\text{w}}$, $a_{\text{s},\text{w}}$,~$b_{\text{s},\text{w}}$, and~$\gamma$ are in general not independent and show a complicated dependence on doping and/or temperature, and on the mean free path. These coefficients are presented in Section~\ref{sec:Aplication} for the case of cuprates and iron-based pnictides.

The free energy is written in the Ginzburg--Landau form of Eq.~\eqref{eq:free_energy} in the vicinity of a critical temperature. However, the critical temperatures~$T_{\text{dw},\text{s}}$ of the transitions into a state with a finite~$W$, respectively,~$\Delta$ may be quite different. We assume that the doping level described by the parameter~$\mu$ is chosen in such a way (${\mu = \mu_{\text{c}}}$), that the critical temperature~$T_{\text{dw}}$ for the non-superconducting OP~$W$ coincides with the critical temperature of the superconducting transition~$T_{\text{s}}$. This is possible because$T_{\text{dw}}$ depends on~$\mu$, whereas~$T_{\text{s}}$ does not. Thus, the coefficients~$a_{\text{s,w}}$ and~$b_{\text{s,w}}$ depend on the differences ${\eta = (1 - T/T_{\text{s}})}$, ${\delta [\mu^{2}] \equiv \mu^{2} - \mu_{\text{c}}^{2}}$, and on impurity concentration~$n_{\text{imp}}$.

The variation of~$\mathcal{F}$ with respect to~$\Delta$ and~$W$ yields the self-consistency (or the Ginzburg--Lanadau) equations
\begin{align}
-\xi_{\text{s}}^{2} \Delta^{\prime \prime} + \Delta \big[ -a_{\text{s}} + b_{\text{s}} \Delta^{2} + \gamma W^{2} \big] &= 0 \,, \label{eq:self_cons_s} \\
-\xi_{\text{w}}^{2} W^{\prime \prime} + W \big[ -a_{\text{w}} + b_{\text{w}} W^{2} + \gamma \Delta^{2} \big] &= 0 \,, \label{eq:self_cons_w}
\end{align}
which represent the foundation of our considerations.

\subsection{Homogeneous case}

Equations~\eqref{eq:self_cons_s} and~\eqref{eq:self_cons_w} without the spatial derivatives yield different uniform solutions, i.e., three different points on the plane of order parameters~$(\Delta, W)$. We denote these points by, respectively, $\Gamma_{\Delta}$ (where ${\Delta \neq 0}$, ${W = 0}$), $\Gamma_{W}$ (where ${\Delta = 0}$, ${W \neq 0}$), and $\Gamma_{W \Delta}$ (where ${\Delta \neq 0}$, ${W \neq 0}$, i.e., both OPs coexist), each corresponding to an extremum of the free energy functional~$\mathcal{F}(\Delta, W)$. Analyzing these points we determine the conditions for a particular point to correspond to a minimum:
\begin{enumerate}
\item $\Gamma_{\Delta}$, where ${\Delta = \sqrt{a_{\text{s}} / b_{\text{s}}}}$, corresponds to a minimum if the second derivatives of~$\mathcal{F}$ with respect to~$\Delta$ and~$W$ are positive, implying
    \begin{align}
    b_{\text{s}} > 0 \,, && \gamma a_{\text{s}} - a_{\text{w}} b_{\text{s}} > 0 \,. \label{eq:ineq_Gd}
    \end{align}
    In particular, the coefficient~$a_{\text{s}}$ must be positive.
\item $\Gamma_{W}$, where ${W = \sqrt{a_{\text{w}} / b_{\text{w}}}}$, corresponds to a minimum if the conditions
    \begin{align}
    b_{\text{w}} > 0 \,, && \gamma a_{\text{w}} - a_{\text{s}} b_{\text{w}} > 0 \,, \label{eq:ineq_Gw}
    \end{align}
    are fulfilled. In particular, the coefficient~$a_{\text{w}}$ must be positive.
\item $\Gamma_{W \Delta}$, where ${\Delta = \sqrt{(a_{\text{s}} b_{\text{w}} - \gamma a_{\text{w}})/ D }}$ and ${W = \sqrt{(a_{\text{w}} b_{\text{s}} - \gamma a_{\text{s}})/ D }}$ with ${D = b_{\text{s}} b_{\text{w}} - \gamma^2}$, corresponds to a minimum provided the conditions
        \begin{align}
    b_{\text{s}} > 0 \,, && b_{\text{w}} > 0 \,, && b_{\text{s}} b_{\text{w}} - \gamma^2 > 0 \,, \label{eq:1st_ineq_Gwd}
    \end{align}
    are satisfied. It follows from the definition of~$D$ and the expressions for~$\Delta$ and~$W$ that the conditions
        \begin{align}
    a_{\text{s}} b_{\text{w}} - \gamma a_{\text{w}} > 0 \,, && a_{\text{w}} b_{\text{s}} - \gamma a_{\text{s}} > 0 \,, \label{eq:2nd_ineq_Gwd}
    \end{align}
    should be fulfilled.
\end{enumerate}
Clearly, the latter inequalities~\eqref{eq:2nd_ineq_Gwd} are incompatible with those in~\eqref{eq:ineq_Gd} and~\eqref{eq:ineq_Gw}. This fact is evident from topological arguments, namely, if the point~$\Gamma_{W \Delta}$ is a minimum of~$\mathcal{F}(\Delta, W)$, then the points~$\Gamma_{\Delta}$ and~$\Gamma_{W}$ can only correspond to a maximum or a saddle point of the free energy functional.

\subsection{Nonhomogeneous case}

In case of heterostructures where the OPs depend on the coordinate~$x$, the most interesting nontrivial solution of the system of equations~\eqref{eq:self_cons_s} and~\eqref{eq:self_cons_w} corresponds to those where the OP~$W$ goes to a finite value~${W_{\infty} = W_{-\infty}}$ (an asymmetric case~${W_{\infty} \neq W_{-\infty}}$ can be considered analogously) while~$\Delta$ vanishes at distances from the interface exceeding~$\xi_{\text{w}}$. In other words, we consider the case when the system is at the point~$\Gamma_{W}$ far away from the interface. We assume that the OP~$W(x)$ is suppressed near the interface, e.g., due to an enhanced impurity scattering in the vicinity of the interface or diffusive scattering on the interface. The diffusive scattering may be caused either by interdiffusion or interface roughnesses. As is known (see~Ref.~\onlinecite{ImpScat} and references within), the critical temperature~$T_{\text{dw}}$ is suppressed by impurity scattering while the critical temperature~$T_{\text{s}}$ of the superconducting transition is only weakly affected. The most essential dependence of the coefficients~$a_{\text{s,w}}$ and $b_{\text{s,w}}$ on impurity scattering is the one of the coefficient~$a_{\text{w}}$.

Doping level near the interface also may be changed due to interdiffusion of atoms. Another reason for a change of the coefficients in the Ginzburg--Landau equations is a different crystal symmetry at the interface. Such a mechanism of enhanced superconductivity at twin boundaries has been considered for conventional low-$T_{\text{c}}$ superconductors in Ref.~\onlinecite{Khlyustikov_Buzdin_87}. A change in the energy spectrum of cuprates near the surface has been calculated in Ref.~\onlinecite{Eschrig_Lankau_Koepernik_2010}. This change can also lead to a modification of coefficients in the Ginzburg--Landau equations or even to a surface superconductivity. It is worth noting that the properties of surface superconductivity in systems with one (superconducting) OP in magnetic field have been studied theoretically in Refs.~\onlinecite{Barzykin_Gorkov_2002,Agterberg_2003,Dimitrova_Feigelman_2007,Agterberg_Babaev_Garaud_2014}.

\begin{figure}[tbp]
\includegraphics[width=1.0\columnwidth]{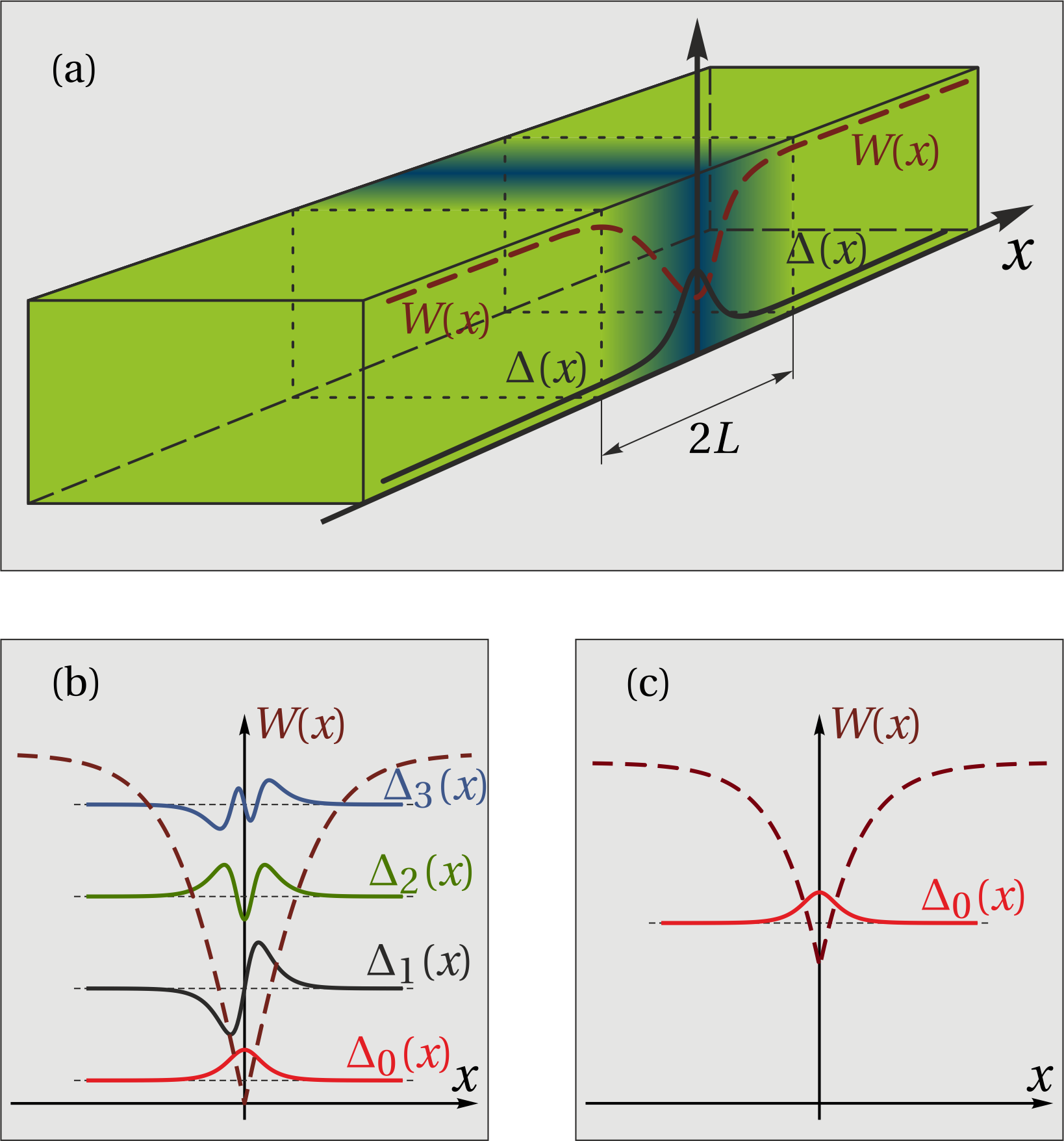}
\caption{(Color online.) (a)~Sketch of considered system. The CDW (or SDW) order parameter~$W$ is suppressed near an interface between two materials in which superconducting order parameter~$\Delta$ may exist alongside~$W$. This suppression leads to the appearance of interfacial superconductivity, which can be thus regarded as `hidden'. (b)~The case of strong suppression of~$W$ with solutions of~$\Delta_n(x)$ given by hypergeometric functions. Note that~$\Delta_0(x)$ has the shape of a soliton, whereas other solutions have nodes. (c)~In the case of a weak suppression of~$W$, one has a shallow `potential' in the `Schr\"{o}dinger' equation and there exists only one `energy' level given by~Eq.~\eqref{eq:single_level}.}
\label{fig:setup}
\end{figure}

The interface behavior of~$W$ can be modeled via the spatial dependence of the coefficient~$a_{\text{w}}$, i.e.,
\begin{align}
a_{\text{w}}(x) = \begin{cases}
                - a_0, & |x| < L \,, \\
                a_{\text{w}}, & |x| > L \,,
                \end{cases}
                \label{eq:a_w}
\end{align}
where~$L$ is a characteristic width of the region where~$W$ is suppressed. The expression for~$a_0$ is presented in Section~\ref{sec:Aplication} for a particular case. Formula for~$a_{\text{w}}(x)$ at ${|x| > L}$ implies that the amplitude~${W_{\pm \infty} = \sqrt{a_{\text{w}} / b_{\text{w}}}}$ is the same at ${x\rightarrow \pm \infty}$, having a lower value at the interface ${x = 0}$, see Fig.~\ref{fig:setup}~(a). We will show that, under these circumstances, a superconducting OP~$\Delta$ arises at the interface decaying to zero as ${x \to \infty}$.

Note that in our previous publication\cite{Moor_Volkov_Efetov_2014} we analyzed nonhomogeneous solutions for the Ginzburg--Landau equations (topological defects) assuming that all the coefficients in these equations are constant. The found solutions may correspond to metastable states with energies higher than that for an uniform solution. In the case considered here, a nonuniform solution for~$\Delta(x)$ is enforced by built-in defect described by the a non-constant coefficient~$a_{\text{s}}(x)$.

In order to find the spatial dependence of~$W(x)$ and~$\Delta(x)$, we assume that the superconducting OP~$\Delta$ is small. Then, in the main approximation the equation for~$W$ acquires the form
\begin{equation}
- \xi_{\text{w}}^2 W^{\prime \prime} + W \big[ - a_{\text{w}}(x) + b_{\text{w}} W^2 \big] = 0 \,,
\end{equation}
where~$a_{\text{w}}(x)$ is given by Eq.~\eqref{eq:a_w}. This equation can be solved exactly, but for simplicity we restrict ourselves to the simplest case of a narrow region of suppression, i.e., ${L \ll \xi_{\text{w}} / \sqrt{a_{0}}}$, thus obtaining the solution
\begin{equation}
W(x) = W_{\infty} \tanh \big[ \kappa_{\text{w}} (|x| + x_0) \big] \,, \label{eq:W(x)}
\end{equation}
where the integration constant~$x_0$ obeys the equation
\begin{equation}
\sinh (2x_{0} \kappa_{\text{w}}) = 4 \xi_{\text{w}}^{2} \kappa_{\text{w}} / a_{0} L \equiv r \,, \label{eq:match_conds}
\end{equation}
with ${\kappa_{\text{w}} = \xi_{\text{w}}^{-1} \sqrt{a_{\text{w}} / 2}}$.

Next, we consider separately the cases of a strong [${r \ll 1}$, cf.~Fig.~\ref{fig:setup}~(b)] and, respectively, weak [${r \gg 1}$, cf.~Fig.~\ref{fig:setup}~(c)] suppression of~$W$ at the interface.

\paragraph*{Strong suppression, ${r \ll 1}$.}
In this case, the product ${2 x_{0} \kappa_{\text{w}} \simeq r}$ is small and the quantity~$x_0$ in Eq.~\eqref{eq:W(x)} can be neglected. Substituting~$W(x)$ in this approximation into Eq.~\eqref{eq:self_cons_s} one obtains
\begin{equation}
\tilde{\xi}_{\text{s}}^2 \Delta^{\prime \prime} + \big[ \mathcal{E} + \mathcal{U} \cosh^{-2} (\kappa_{\text{w}} x) \big] \Delta = g \Delta^3 \,, \label{eq:GP_main}
\end{equation}
where ${\tilde{\xi}_{\text{s}} = \xi_{\text{s}} \sqrt{b_{\text{w}}}}$, ${\mathcal{E} = a_{\text{s}} b_{\text{w}} - \gamma a_{\text{w}}}$, ${\mathcal{U} = \gamma a_{\text{w}}}$, and ${g = b_{\text{s}} b_{\text{w}}}$. Equation~\ref{eq:GP_main} has a form of the Gross--Pitaevskii equation.\cite{Gross,Pitaevskii} Solving this equation one can determine the spatial dependence of the superconducting OP~$\Delta$. The calculations in this case formally coincide with those carried out in Ref.~\onlinecite{Moor_Volkov_Efetov_2014} if the interchange ${\Delta \leftrightarrow W}$ is done. For completeness we repeat the steps one has to perform seeking for a solution~$\Delta(x)$. Assuming that~$\Delta$ is small, the right-hand side of Eq.~\eqref{eq:GP_main} can be neglected and we are left with the linearized Gross--Pitaevskii equation, i.e., the `Schr\"{o}dinger' equation, which can be solved considering the eigenvalue problem
\begin{equation}
\hat{\mathcal{L}} \Delta_n = \mathcal{E}_n \Delta_n \,,
\end{equation}
where the operator~${\hat{\mathcal{L}} = - \tilde{\xi}_{\text{s}}^2 \partial_{xx}^2 - \mathcal{U} \cosh^{-2} (\kappa_{\text{w}} x)}$, and~$\Delta_n$ and~$\mathcal{E}_n$ are the eigenfunctions and eigenvalues of~$\hat{\mathcal{L}}$. The solutions~$\Delta_n$ corresponding to a discrete spectrum of~$\mathcal{E}_n$ can be expressed in terms of hypergeometric functions and the `energy' levels (${\mathcal{E} < 0}$) being given by~\cite{LLquantMech}
\begin{equation}
\mathcal{E}_{n} = - \frac{\tilde{\xi}_{\text{s}}^{2} \kappa_{\text{w}}^{2}}{4} \Bigg[ - (1 + 2 n) + \sqrt{ 1 + \frac{4 \mathcal{U}}{\tilde{\xi}_{\text{s}}^{2} \kappa_{\text{w}}^{2}}} \Bigg]^{2} \,. \label{eq:eigenvalues}
\end{equation}
Provided the inequality~${4 \mathcal{U}/\tilde{\xi}_{\text{s}}^{2} \kappa_{\text{w}}^{2} < 8}$ is fulfilled, there is only one `energy' level with ${n = 0}$ and the corresponding solution~$\Delta_0(x)$ has a form of a soliton. Otherwise there are several solutions decaying far away from the interface and corresponding to~$\mathcal{E}_n$.

Representing the OP~$\Delta$ close to a certain `energy' level~$\mathcal{E}_n$ as ${\Delta(x) = c_n \Delta_n(x) + \delta \Delta_n(x)}$ with a small correction~$\delta \Delta_n(x)$ orthogonal to~$\Delta_n(x)$, one obtains the coefficients~$c_n$,
\begin{equation}
c_n^2 = \frac{\mathcal{E} - \mathcal{E}_n}{g \langle\langle \Delta_n^4(x) \rangle\rangle} \,, \label{eq:c_n}
\end{equation}
where $\langle\langle f(x) \rangle\rangle = \int_{-\infty }^{\infty} \mathrm{d} x \, f(x)$ (double angle brackets are used to distinguish the notation from the averaging over momenta directions introduced in Appendix~\ref{sec:doping_dependence}).

Note an important point. The condition ${\mathcal{E} < 0}$ that determines the appearance of the interface superconductivity coincides with the condition~\eqref{eq:ineq_Gw} that provides the stability of the state with ${W \neq 0}$ and ${\Delta = 0}$ in the bulk. This means that if a nonsuperconducting state in the bulk is characterized by a nonzero OP~$W$, any suppression of~$W$ leads to the appearance of local superconductivity. We demonstrate this considering the case of a small suppression of~$W$.

\begin{figure}[tbp]
\includegraphics[width=0.7\columnwidth]{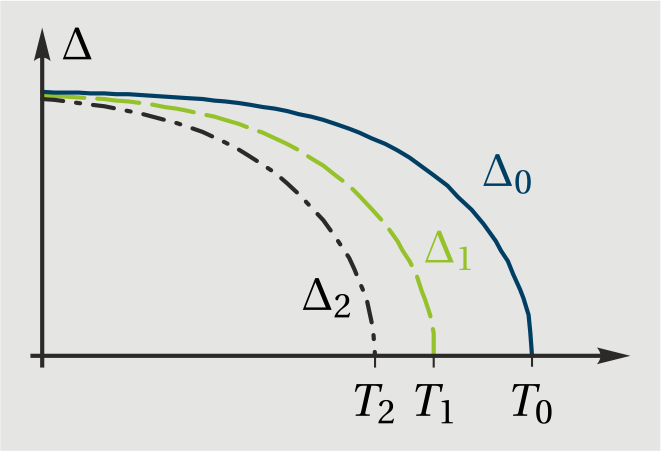}
\caption{(Color online.) Sketch of temperature dependence of the amplitudes~$\Delta_{n}$ corresponding to different eigenvalues~$\mathcal{E}_{n}$, see Eq.~\eqref{eq:eigenvalues}. At a given temperature~$T$, the state with the largest~$\Delta_n$, i.e., the state~$\Delta_0$, corresponds to a minimum of the free energy. However, the transitions between different~$\Delta_n$ are possible}
\label{fig:Delta_n}
\end{figure}

\paragraph*{Weak suppression, ${r \gg 1}$.} In this case,  one obtains from Eq.~\eqref{eq:match_conds} ${x_0 \kappa_{\text{w}} \simeq \ln \sqrt{2 r}}$ and the spatial dependence of~$\Delta(x)$ is determined by the `Schr\"{o}dinger' equation.~\eqref{eq:GP_main} with the `potential' ${\mathcal{U} \cosh^{-2}(\kappa_{\text{w}} x) \to \mathcal{V}(x) = 1 - \tanh^2 \big[ \kappa_{\text{w}} (|x| + x_0) \big]}$. In other words, the function~$\Delta(x)$ is determined by the `Schr\"{o}dinger' equation that provides the `energy' levels in a shallow potential well~$\mathcal{V}(x)$. As is well known,\cite{LLquantMech} there always exists a single `energy' level
\begin{equation}
\mathcal{E}_0 = - J^2 / (2 \tilde{\xi}_{\text{s}})^2 \,,
\label{eq:single_level}
\end{equation}
where ${J = \langle\langle \mathcal{V}(x) \rangle\rangle = 4 \kappa_{\text{w}}^{-1} \exp(-2 \kappa_{\text{w}} x_0 )}$. The amplitude~$c_{0}$ is given again by Eq.~\eqref{eq:c_n} with ${n = 0}$. This means that a superconducting condensate with a small amplitude ${\Delta \simeq T_{\text{s}} \mathcal{E}_{0}}$, where~$T_{\text{s}}$ is the superconducting transition temperature in the bulk, necessarily arises at the interface as soon as the competing OP is arbitrarily weakly suppressed.

Note that, in the case of a two-dimensional point defect instead of the interface,~$|\mathcal{E}_{0}|$ is an exponentially small\cite{LLquantMech} quantity and, therefore, the radius of the decay of the condensate is exponentially large.

\section{Application to cuprates and pnictides}
\label{sec:Aplication}

\subsection{Relation of coefficients to microscopic parameters}

Here, we present the expressions for the coefficients in the Ginzburg--Landau expansion for the case of cuprates and iron-based pnictides.

As has been shown in Ref.~\onlinecite{MVVE14}, the model that has been developed in detail in Refs.~\onlinecite{Chubukov10,Chubukov10a,Schmalian10} for Fe\nobreakdash-based pnictides (generally, to two-band superconductors with an SDW), is applicable to quasi-one-dimensional superconductors with a CDW, and, after certain modification, also to cuprates.

First, we consider the region ${|x| > L}$ and assume that the impurity concentration in this region is small, i.e., the mean free path ${l \gg \xi_{\text{s,w}}}$. In the model of Refs.~\onlinecite{Chubukov10,Chubukov10a,Schmalian10}, the coefficients are related to the microscopic parameters of the model via ${a_{\text{s}} = \eta}$, ${b_{\text{s}} \simeq 1.05}$, ${a_{\text{w}} = \eta (1 - \beta_{1}) - \langle \beta_{2} \delta [\mu^{2}] \rangle }$, ${b_{\text{w}} = s_{3m}}$, ${\gamma = s_{2m}}$, where ${\eta = 1 - T/T_{\text{s}}}$ and ${\delta [\mu^{2}] = \mu^2 - \mu_{\text{c}}^2}$ with~$\mu$ being a function that describes the curvature of the Fermi sheets of the quasi-one-dimensional superconductor or a deviation of the average Fermi surface from the perfect circle in iron-based superconductors, where partial nesting between the elliptical electron bands and the circular hole bands leads to the formation of the spin-density wave. The functions~$s_{2m}$, $s_{3m}$ and $\beta_{1,2}$ (see Appendix~\ref{sec:doping_dependence}) depend on the dimensionless critical curvature ${m = \mu_{\text{c}} / \pi T_{\text{s}}}$ defined in such a way that the critical temperature~$T_{\text{dw}}$ of the formation of the OP~$W$ equals~$T_{\text{s}}$, where $T_{\text{s},\text{dw}}$ are the critical temperatures for the transition into the, correspondingly, superconducting and CDW or SDW state in absence of the competing order and doping. The critical~$\mu_{\text{c}}$ is determined by the equation (see Appendix~\ref{sec:doping_dependence})
\begin{equation}
\langle 2 \mu_{\text{c}}^2 s_{1m} (\mu_{\text{c}}) \rangle = \ln (T_{\text{dw}} / T_{\text{s}}) \,, \label{mu_c}
\end{equation}
where the critical temperature~$T_{\text{dw}}$ depends, generally speaking, on impurity concentration which is assumed to be small far from the interface.

Next, consider the region ${|x| < L}$, where the impurity scattering is assumed to be stronger. In this case, the temperature ${T_{\text{dw}} \simeq T_{\text{dw0}} \big[ 1 - (4 \pi T_{\text{dw0}} \tau)^{-1} \big]}$, where~$T_{\text{dw0}}$ is the critical temperature of the transition into a state with~${W \neq 0}$ in the absence of impurities and superconductivity, ${\tau = l / v}$ is the momentum relaxation time for intraband scattering, and the mean free path~$l$ is assumed to be larger than~$v / T_{\text{dw0}}$. One can easily show that, in this case, ${a_{0} = a_{\text{w}} - (4 \pi T_{\text{dw0}} \tau)^{-1}}$, where~$a_{\text{w}}$ is given by the expression above. Provided the condition ${a_{\text{w}} \ll (4 \pi T_{\text{dw0}} \tau)^{-1} \ll 1}$ is fulfilled, then the coefficient~$a_{0}$ in Eq.~\eqref{eq:a_w} is positive (meaning that ${-a_{0} < 0}$) and considerably esceeds~$a_{\text{w}}$.

\subsection{Temperature and doping dependence}

Considering the expression~\eqref{eq:c_n} for the coefficients~$c_n$, one can see that at small difference ${|\mathcal{E} - \mathcal{E}_{n}|}$, the OP~$\Delta $ is also small. It turns to zero at certain temperatures or doping levels, where ${\mathcal{E}(T_{n}, \mu_{m}) = \mathcal{E}_{n}(T_{n}, \mu_{n})}$ holds, with~$\mathcal{E}_{n}(T_{n}, \mu_{n})$ given by Eq.~\eqref{eq:eigenvalues} with coefficients expressed through the microscopic parameters and the temperature. In Fig.~\ref{fig:Delta_n} we plot the temperature dependence of the amplitudes~$\Delta_{n}$ corresponding to different eigenvalues~$\mathcal{E}_{n}$. Clearly, when the temperature~$T$ becomes lower than the temperature~$T_{0}$ determined by Eq.~\eqref{eq:eigenvalues}, the superconducting OP~$\Delta_{0}(x)$ arises at the interface with the amplitude increasing when~$T$ is lowered. The temperature~$T_{0}$ is lower than~$T_{\text{s}}$ (${\eta > 0}$), but, under certain conditions, higher than the temperature~$T_{\text{b}}$ at which the superconducting state becomes more favorable in the bulk. At ${T < T_{1}}$, a new branch~$\Delta_{1}(T)$ appears, etc.

The minimal temperature~$T_{n}$ (or maximal $n_{\text{max}}$) at which the branch~$\Delta_{n_{\text{max}}}(T)$ appears, is determined by the condition ${2 n_{\text{max}} \leq \sqrt{1 + 4 \mathcal{U} \tilde{\xi}_{\text{s}}^{-2} \kappa_{\text{w}}^{-2}} - 1}$. It can be shown that at a given temperature~$T$, the state with the largest~$\Delta_{n}$, i.e., the state~$\Delta_{0}$, corresponds to a minimum of the free energy. However, the transitions between different~$\Delta_{n}$ are possible analogously to transitions between an overcooled state and equilibrium.

\subsection{Interface superconducting transition temperature}

Our considerations concerned the case when the system is at the point~$\Gamma_{W}$ far away from the interface, i.e., the conditions~\eqref{eq:ineq_Gw} are valid while the conditions~\eqref{eq:ineq_Gd} are violated. When applied to the case of quasi one-dimensional materials with the CDW or to material like iron-based pnictides with an SDW, these conditions can be presented in the form
\begin{align}
A_{2} \eta + B_{2} \delta [\mu^{2}] < 0 \,, && A_{1} \eta + B_{1} \delta[\mu^{2}] < 0 \,, && s_{3m} > 0 \,,
\end{align}
where ${A_{1} = s_{3m} - s_{2m} (1 - \beta_{1})}$, ${B_{1} = s_{2m} \beta_{2}}$, ${A_{2} = s_{2m} - s_{3}(1 - \beta_{1})}$, and ${B_{2} = 1.05 \beta_{2}}$ if expressed via the microscopic parameters of the model. The coefficients~$A_{1,2}$ and $B_{1,2}$ depend on~$\mu(\mu_{0}, \mu_{\varphi})$ and may attain positive or negative values (only~$B_{2}$ is a positive quantity). If these coefficients are all positive, then this conditions can be fulfilled provided ${\delta[\mu^{2}] < 0}$. Thus, we can rewrite them as ${B_{2} |\delta[\mu^{2}]| > A_{2} \eta}$ and ${B_{1}|\delta[\mu^{2}]| > A_{1} \eta}$. In terms of microscopic parameters the latter can be written as
\begin{align}
\eta < c_{\mu} B_{1} / A_{1} \equiv 1 - \tilde{T}_{W} \,, && \eta < c_{\mu} B_{2} / A_{2} \equiv 1 - \tilde{T}_{\Delta} \,,  \label{eq:conds_microscopic}
\end{align}
where $c_{\mu} = |\delta[\mu^{2}]|$ and we introduced the `critical' temperatures~${\tilde{T}_{W, \Delta} = T_{W, \Delta} / (\pi T_{\text{dw},\text{s}})}$.

\begin{figure}[tbp]
\includegraphics[width=1.0\columnwidth]{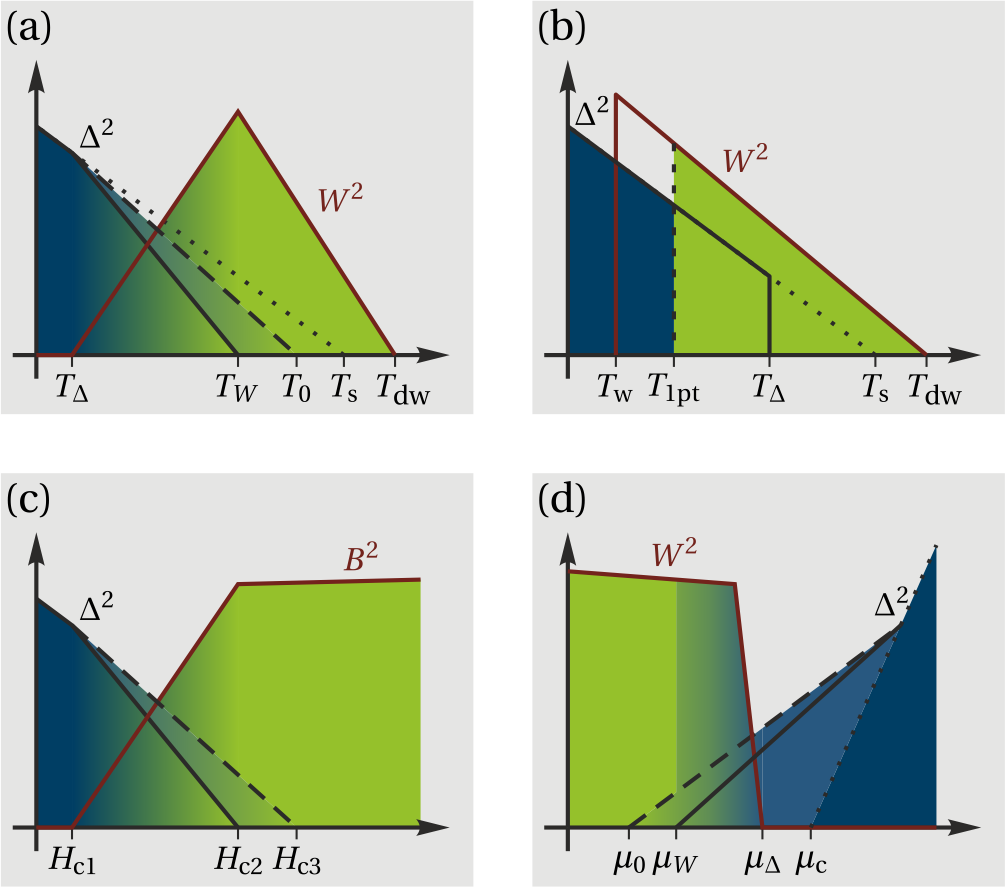}
\caption{(Color online.) (a)~In the case ${T_{\Delta} < T_{W}}$, three ranges of temperature. For ${T < T_{\Delta}}$, the system is in a pure superconducting state, whereas in the temperature range ${T_{\Delta } < T < T_{W}}$, a mixed state (the state of coexistence) with ${\Delta \neq 0}$ and ${W \neq 0}$  is realized. Thus, the bulk superconducting transition temperature corresponds to~$T_{W}$. The temperature range of the coexistence state is widened (enhancement of superconductivity) on appearance of the interface superconductivity with the highest transition temperature~$T_0$ (there may be other transition temperatures~$T_n$). Finally, for ${T > T_0}$, the system is in a pure $W$-state (CDW or SDW) up to its transition into the normal state at~$T_{\text{dw}}$. (b)~In case ${T_{\Delta} > T_{W}}$, the system may be in either the pure superconducting or in the pure $W$-state. The transition from one into another is of the first order at a temperature~$T_{1\text{pt}}$. If~$T_0$ falls into the region ${T_{1\text{pt}} < T_{\text{s}}}$, which is possible if the difference ${T_{\text{s}} - T_{1\text{pt}}}$ is positive and sufficiently large, then, at the interface, superconductivity is induced. (c)~An analogy with conventional second-type superconductors in a magnetic field~$H$ is sketched. The quantities~$T_{\Delta, W}$ and~$T_0$ are analogous to the critical fields~$H_{\text{c}1,\text{c}2}$ and~$H_{\text{c}3}$, respectively, i.e., up to~$H_{\text{c}1}$ the system is in a purely superconducting state (full expulsion of the magnetic field), in a mixed state in the interval ${H_{\text{c}1} < H < H_{\text{c}2}}$, and in the Abrikosov's vortex state for ${H_{\text{c}2} < H < H_{\text{c}3}}$ loosing the superconducting properties for ${H > H_{\text{c}3}}$. (d)~If, instead of temperature, the doping (or curvature)~$\mu$ is considered, the situation is similar to the temperature dependence.}
\label{fig:temperature_dependence}
\end{figure}

One can distinguish two from the physical point of view different cases:
\begin{itemize}
\item[a)] The case ${T_{\Delta} < T_{W}}$ is realized if the quantity ${D \equiv A_{1} B_{2} - A_{2} B_{1} = s_{3} s_{3m} - s_{2m}^{2} > 0}$ provided that~$A_1$ and~$A_2$ are positive. In this case, the pure superconducting state (minimum of the free energy at~$\Gamma_{\Delta}$) exists at temperatures ${T < T_{\Delta}}$ with $\Delta_{\text{un}}^{2} = a_{\text{s}} / b_{\text{s}}$. In the temperature range ${T_{\Delta } < T < T_{W}}$, a mixed state (the state of coexistence) with ${\Delta \neq 0}$ and ${W \neq 0}$ given by expressions just before Eq.~\eqref{eq:1st_ineq_Gwd} takes place. At ${T > T_{0}}$, a pure CDW-state or, more generally, a W\nobreakdash-state occurs with $W_{\text{un}}^{2} = a_{\text{w}} / b_{\text{w}}$. In the interval ${T_{W} < T < T_{0}}$, one has a surface (or interface superconductivity), where~$T_{0}$ is the temperature determined by Eq.~\eqref{eq:eigenvalues} corresponding to ${n = 0}$ (the ground state). At ${T_{0} < T < T_{\text{dw}}}$, the system is nonsuperconducting with the OP~${W \neq 0}$. In Fig.~\ref{fig:temperature_dependence}~(a) we show schematically the temperature dependence of~$\Delta$ and~$W$ and also the temperature range in which the local superconductivity exists. At temperatures~$T_{W, \Delta}$, second-order phase transitions occur and the OPs~$\Delta$ and~$W$ arise continuously (see expressions for~$\Delta$ and~$W$ before Eq.~\eqref{eq:1st_ineq_Gwd} where ${\Delta \sim \sqrt{T_{W} - T}}$ and ${W \sim \sqrt{T - T_{\Delta}}}$). Note an obvious analogy with conventional second-type superconductors in a magnetic field~$H$.\cite{Abrikosov_book,deGennes} The quantities~$T_{\Delta, W}$ are analogous to the critical fields~$H_{\text{c}1,\text{c}2}$ [cf.~Fig.~\ref{fig:temperature_dependence}~(c)] so that at ${T < T_{W}}$ one has a purely superconducting state (full expulsion of the magnetic field), a mixed state in the interval ${T_{W} < T < T_{\Delta}}$ (correspondingly, the Abrikosov's vortex state) and a surface or interface superconductivity in the range ${T_{\Delta} < T < T_{0}}$ (correspondingly, in the range ${H_{\text{c}2} < H < H_{\text{c}3}}$). At last, at ${T > T_{0}}$, one has a pure $W$\nobreakdash-state which corresponds to the normal state in conventional superconductors. Note that same analysis can be performed if, instead of temperature, the doping (or curvature) given by~$c_{\mu}$ in Eq.~\eqref{eq:conds_microscopic} is considered and one obtains the corresponding situation as depicted in Fig.~\ref{fig:temperature_dependence}~(d).
\item[b)] The case ${T_{\Delta} > T_{W}}$ is realized if the quantity ${D \equiv A_{1} B_{2} - A_{2} B_{1} = s_{3} s_{3m} - s_{2m}^{2} < 0}$ provided that~$A_1$ and~$A_2$ are positive. In this case, the pure superconducting state (minimum of the free energy at~$\Gamma_{\Delta}$) exists again at temperatures ${T < T_{\Delta}}$, but this minimum is global only at ${T < T_{1\text{pt}}}$, where the critical temperature~${T_{1\text{pt}}}$ for the first-order phase transition is determined by the equation ${\mathcal{F}(a_{\text{w}} / b_{\text{w}}) = \mathcal{F}(a_{\text{s}} / b_{\text{s}})}$. At ${T > T_{\text{dw}}}$, a uniform solution for ${W = \sqrt{a_{\text{w}} / b_{\text{w}}}}$ arises, but it corresponds to a global minimum at ${T > T_{1\text{pt}}}$. In this case, no region of coexistence exists, and at ${T = T_{1\text{pt}}}$ a first-order phase transition from the superconducting state to a state with ${W \neq 0}$ takes place with increasing temperature, see Fig.~\ref{fig:temperature_dependence}~(b). Now, if~$T_0$ as determined by Eq.~\eqref{eq:eigenvalues} corresponding to ${n = 0}$ (the ground state) falls into the region ${T_{1\text{pt}} < T_{\text{s}}}$, which is possible if the difference ${T_{\text{s}} - T_{1\text{pt}}}$ is positive and sufficiently large, then, at the interface, superconductivity is induced.
\end{itemize}

\subsection{Experiments}

The obtained appearance of superconductivity (or enhancement of the critical transition temperature) at an interface between two materials in which, alongside superconductivity, another exists and is energetically more favorable, may be realized in two prominent examples of such systems. One of them, the cuprates, show a charge-density wave order alongside superconductivity\cite{Lawler_et_al_2010,Parker_et_al_2010,Julien_2011,Julien_2013,Ghiringhelli_et_al_2012,Chang_et_al_2012,Achkar_et_al_2012,LeBoeuf_et_al_2013,Blackburn_et_al_2013,Comin_et_al_2014,Fujita_et_al_2014} In these materials, an enhancement of superconducting transition temperature has been found in a bilayer constructed of La$_{1.65}$Sr$_{0.35}$CuO$_{4}$ and La$_{1.875}$Ba$_{0.125}$CuO$_{4}$.\cite{Yuli_et_al_2008,Koren_Milo_2010}

Superconductivity accompanied by a spin-density wave is known to exist in the iron-based pnictides, where the interface superconductivity is proposed to be the driving effect behind the almost doubling of superconducting transition temperature in CaFe$_2$As$_2$.\cite{Reich_2013,Chu_arXiv_2013}

Unfortunately, there are no data on spatial dependence of the order parameter accompanying superconductivity in these experiments, neither the spatial dependence of the superconducting order parameters has been investigated in these experiments. Such a measurement would provide a test of our theory, if charge- or spin-density wave would have been suppressed near the interface.

Another interesting effect also serving as a test of our predictions is related to the fact that there might appear a hysteretic behavior stemming from the presence of different `energy' levels [see Eq.~\eqref{eq:eigenvalues}]. This results in a sequence of temperatures~$T_n$ at which~$\Delta$ formally vanishes but the temperature dependence of~$\Delta$ is determined by the highest temperature of them all, i.e., by~$T_0$ since the minimum of the free energy is deepest here. Interestingly, at a temperature~$T_n$ a local minimum of the free energy is present and adjusting~$\Delta$ by means, e.g., of an external field, one can let it follow the temperature dependence of~$\Delta_n$ after the relaxation of external constrains, so, indeed, ``there's a lot of room for new combinations''.\cite{Reich_2013}

\section{Discussion}
\label{sec:Discussion}

We have studied a system with two competing OPs one of which is the superconducting OP~$\Delta$ and another,~$W$, can be the amplitude of the charge- or spin-density wave. On the basis of Ginzburg--Landau equations we have shown that if the temperature and doping are chosen in such a way that the state with ${W \neq 0}$ and ${\Delta = 0}$ is favorable, i.e., it corresponds to a minimum of the free energy in the bulk of the sample, an arbitrary small suppression of~$W$ at an interface or a defect leads to the appearance of local superconductivity. This mechanism of local superconductivity in a system with two OPs may be responsible for the interface superconductivity observed in many materials including cuprates and iron-based pnictides. As is firmly established, in cuprates and iron-based pnictides, CDW (or quadrupole charge order~\cite{Pseudogap}) or SDW can exist alongside with superconductivity.

We found that in case of a strong suppression of~$W$ at some point, there are several solutions for the superconducting OP~$\Delta(x)$ which are localized on the scale of the coherence length~$\xi_{\text{s}}$. These solutions are found from the nonlinear Schr\"{o}dinger equation (or Gross--Pitaevskii equation) and correspond to different `energy' levels. Each solution arises at certain values of temperatures~$T_{n}$ (or doping level~$\mu_{n}$) and has a different form changing from a soliton-like one to an oscillatory function decaying at infinity. However, only the soliton-like solution~$\Delta_{0}(x)$ corresponds to a minimum of the free energy. Other solutions~$\Delta_{n}(x)$ (with ${n \neq 0}$) with nodes have higher energies. They correspond to metastable states. If the corresponding potential~$\mathcal{V}(x)$ in the Schr\"{o}dinger equation is not deep enough, there is only one `energy' level and only one soliton-like solution for~$\Delta(x)$.

In the case of an asymmetric potential $\mathcal{V}(x)$, the localized solution for~$\Delta(x)$ exists provided that the potential well is deep enough.\cite{LLquantMech} If for some reasons the OP~$W$ is suppressed at the surface, the solutions for~$\Delta(x)$ have the same form as in case of the symmetric~$\mathcal{V}(x)$ and surface superconductivity arises in the sample. Note an important point. The local superconductivity may arise in one- or two-dimensional cases (a flat interface or a point defect) if the suppression of the OP~$W$ is small. This means that even if the minimal value of~$W(x)$ corresponds to a minimum of the free energy in an uniform case, in a nonuniform case local superconductivity would arise at the interface or at the surface.

The developed theory is able to explain the emergent or enhanced interface superconductivity in some cuprate or iron-based pnictide heterostructures. The predictions can be tested in further experiments.

Remarkable, the used approach based on the consideration of Ginzburg--Landau equations for two coupled order parameters does not depend on the nature of order described by those. The obtained results are also applicable to description of an arbitrary `hidden' order evolving at the interface, e.g., the ferromagnetism induced at the interface of an oxide heterostructure observed recently.\cite{Salluzzo_et_al_2013}

Note that localized superconductivity may arise also in a homogeneous sample if a nonuniform~$W(x)$ state (for example, stripes) is energetically favorable due to an internal mechanism (e.g., the Larkin--Ovchinnikov--Fulde--Ferrel mechanism).\cite{Larkin_Ovchinnikov_1965,Fulde_Ferrell_1964} Analysis of this state with two competing order parameters deserves a separate consideration.\footnote{The possibilitiy of such a state with two coexisting order parameters was noted in Refs.~\onlinecite{Agterberg08,Gorkov_13,Chubukov10,Schmalian10}.}

\begin{acknowledgments}
We appreciate the financial support from the DFG via the Projekt~EF~11/8\nobreakdash-1; K.~B.~E.~gratefully acknowledges the financial support of the Ministry of Education and Science of the Russian Federation in the framework of Increase Competitiveness Program of NUST~``MISiS'' (Nr.~K2-2014-015). We thank Prof.~Dr.~I.~Eremin for useful comments and valuable discussions.
\end{acknowledgments}

\appendix

\section{Doping dependence of the coefficients in Ginzburg--Landau equations}
\label{sec:doping_dependence}

In the notation of Refs.~\onlinecite{Chubukov10,Chubukov10a,Schmalian10}, the Ginzburg--Landau equations have the form
\begin{align}
-\xi_{\text{s}}^{2} \nabla^{2} \Delta + \Delta \big[ W^{2} s_{2m} + \Delta^{2} s_{3} - \ln(T_{\text{s}}/T) \big] &= 0 \,, \label{1} \\
-\xi_{\text{w}}^{2} \nabla^{2} W + W \big[ \langle 2 \mu^{2} s_{1m} \rangle + W^{2} s_{3m} + \Delta^{2} s_{2m} - \ln(T_{\text{dw}}/T) \big] &= 0 \,, \label{1'}
\end{align}
where~$\xi_{\text{s},\text{w}}$ are the coherence lengths (at low temperatures) for~$\Delta$ and~$W$, respectively, and~$T_{\text{s},\text{dw}}$ are, respectively, the critical temperatures for the transition into the pure superconducting state or into a state with a CDW or an SDW only. In other words,~$T_{\text{dw}}$ is the critical temperature for the transition into the charge-ordered state in absence of~$\Delta$ and~$\mu$, while~$T_{\text{s}}$ is the superconducting transition temperature in absence of~$W$. The angle brackets mean the angle averaging (in Fe\nobreakdash-based pnictides) or integration along the sheets of the Fermi surfaces in quasi-one-dimensional superconductors. The functions~$s_{1m}$,~$s_{2m}$, etc.~are functions of the normalized curvature ${m=\mu /(\pi T_{s})}$, where ${\mu = \mu_{0} + \mu_{\varphi} \cos \big[(p_{y}^{2} + p_{z}^{2})^{1/2} a \big]}$ is a curvature in quasi-one-dimensional superconductors with a doping-dependent value of~$\mu_{0}$. It is assumed that the Fermi surface of these superconductors consists of two slightly curved sheets which are perpendicular to the~$x$~axis~\cite{MVVE14}. In the case of Fe\nobreakdash-based pnictides, ${\mu = \mu_{0} + \mu_{\varphi} \cos (2\varphi)}$ is a quantity that describes an elliptic (${\mu_{\varphi} \neq 0}$) and circular (${\mu_{\varphi} = 0}$) Fermi surfaces of electron and hole bands.\cite{Chubukov10,Chubukov10a,Schmalian10} All quantities---$\Delta$,~$W$ and~$\mu $---are measured in units of~$\pi T_{\text{s}}$. The expressions for the coefficients in the G\nobreakdash--L~expansion with account for impurity scattering have been calculated in Ref.~\onlinecite{Syzranov14}.

Replacing the derivative ${\nabla \to \nabla - \mathrm{i} 2 \pi A / \Phi_{0}}$, one can use Eqs.~\eqref{1} and~\eqref{1'} to describe vortices in superconductors with a CDW,\cite{Efetov13a} where~$\Phi_{0}$ is the magnetic flux quantum.

As it is seen from Eq.~\eqref{1'}, the critical temperature~$T_{\text{dw}}$ depends on doping, i.e., on the parameter~$\mu$. We choose this parameter ${\mu = \mu_{\text{c}}}$ in such a way that ${T_{\text{dw}}(\mu_{\text{c}}) = T_{\text{s}}}$. This means that at ${T = T_{\text{s}}}$, the quantities ${\Delta = W = 0}$, and, thus,~$\mu_{\text{c}}$ obeys the equation
\begin{equation}
\langle 2 \mu_{\text{c}}^{2} s_{1m}(\mu_{\text{c}}) \rangle = \ln ( T_{\text{dw}}/T_{\text{s}} ) \,, \label{2}
\end{equation}
where~$\mu_{\text{c}}$ is a function of two parameters, i.e.,~${\mu_{\text{c}} = \mu_{\text{c}}(\mu_{0}, \mu_{\varphi}})$.

Then, we expand the function~$s_{1m}(\mu ,T)$ in the deviations ${\delta [\mu^{2}] = \mu^{2} - \mu_{\text{c}}^{2}}$ and ${\delta T = T_{\text{s}} - T}$, thus obtaining ${s_{1m}(\mu, T) = s_{1m}(\mu_{\text{c}}, T_{\text{s}}) + \beta_{1} \delta T + \langle \beta_{2} \delta [\mu^{2}] \rangle}$, and use Eq.~(\ref{2}) to obtain equations in a general standard form (assuming that all the functions depend only on one coordinate~$x$),
\begin{align}
-\xi_{\text{s}}^{2} \Delta^{\prime \prime} + \Delta \big[ -a_{\text{s}} + b_{\text{s}} \Delta^{2} + \gamma W^{2} \big] &= 0 \,, \label{3} \\
-\xi_{\text{w}}^{2} W^{\prime \prime} + W \big[ -a_{\text{w}} + b_{\text{w}} W^{2} + \gamma \Delta^{2} \big] &= 0 \,, \label{3'}
\end{align}
with~$\Delta^{\prime}$ and~$W^{\prime}$ as well as~$\Delta^{\prime \prime}$ and~$W^{\prime \prime}$ denoting the first and second derivatives with respect to~$x$, respectively. These equations determine extrema of the free energy functional [cf.~Eq.~\eqref{eq:free_energy}]
\begin{align}
\mathcal{F} = \frac{1}{2} \int \mathrm{d} x & \Bigg[ \xi_{\text{s}}^{2} (\Delta^{\prime})^{2} - a_{\text{s}} \Delta^{2} + \frac{b_{\text{s}}}{2} \Delta^{4} + \gamma \Delta^{2} W^{2} + \\
& + \xi_{\text{w}}^{2} (W^{\prime})^{2} - a_{\text{w}} W^{2} + \frac{b_{\text{w}}}{2} W^{4} \Bigg] \,, \notag
\end{align}
with respect to~$\Delta$ and~$W$, and the corresponding coefficients of the G\nobreakdash--L~expansion are related to variables in Eqs.~(\ref{1}) and~(\ref{1'}) via ${a_{\text{s}} = \eta}$, ${b_{\text{s}} = s_{3} \simeq 1.05}$, ${a_{\text{w}} = \eta (1 - \beta_{1}) - \langle \beta_{2} \delta [\mu^{2}] \rangle }$, ${b_{\text{w}} = s_{3m}}$, ${\gamma = s_{2m}}$, where ${\eta = 1 - T/T_{\text{s}}}$. The expressions for the coefficients in terms of the microscopic parameters of the model for cuprates and iron-based pnictides are given as follows:
\begin{align}
s_3 &= \sum_{n=0}^{\infty}
(2 n + 1)^{-3} \,, \\
s_{1m} &= \sum_{n=0}^{\infty}
(2 n + 1)^{-1} \big[ (2 n + 1)^2 t^2 + m^2 \big]^{-1} \,, \\
s_{2m} &= \sum_{n=0}^{\infty}
\big\langle \big[ (2 n + 1)^2 - m^2 \big] (2 n + 1)^{-1} \big[ (2 n + 1)^2 + m^2 \big]^{-2} \big\rangle \,, \\
s_{3m} &= \sum_{n=0}^{\infty}
\big\langle (2 n + 1) \big[ (2 n + 1)^2 - 3 m^2 \big] \big[ (2 n + 1)^2 + m^2 \big]^{-3} \big\rangle \,, \\
\beta_1 &= \sum_{n=0}^{\infty}
\big\langle 4 m^2 (2 n + 1) \big[ (2 n + 1)^2 + m^2 \big]^{-2} \big\rangle \,, \\
\beta_2 &= \sum_{n=0}^{\infty}
2 (2 n + 1)^{-1} \big[ (2 n + 1)^2 + m^2 \big]^{-1} \,,
\end{align}
where ${t = T/T_{\text{s}}}$, the angle brackets~$\langle \ldots \rangle$ denote the angle averaging (in iron-based pnictides) or integration along the sheets of the Fermi surfaces (in quasi-one-dimensional superconductors or cuprates).

\section{Details on solution of the Gross--Pitaevskii equation}
\label{sec:solution_of_GP}

Here we sketch the solution of the Gross-Pitaevskii equation for~$\Delta$. In zero-order approximation we obtain for~$\Delta_{0}$ from Eq.~\eqref{eq:GP_main}
\begin{equation}
\tilde{\xi}_{\Delta}^{2} \Delta_{0}^{\prime \prime} + \Delta_{0} \big[ \mathcal{E} + \mathcal{U} \cosh^{-2}(\kappa_{\text{w}} x) \big] =0 \,. \label{eq:zero_order_delta}
\end{equation}
This equation is integrable and its solutions~$\psi_{n}$ corresponding to a discrete spectrum of~$\mathcal{E}_{n}$ are expressed in terms of hypergeometric functions~\cite{LLquantMech}. In our notations, the `energy' levels of discrete spectrum are given by~\cite{LLquantMech}
\begin{equation}
\mathcal{E}_{n} = - \frac{\tilde{\xi}_{\text{s}}^{2} \kappa_{\text{w}}^{2}}{4} \Bigg[ - (1 + 2 n) + \sqrt{ 1 + \frac{4 \mathcal{U}}{\tilde{\xi}_{\text{s}}^{2} \kappa_{\text{w}}^{2}}} \Bigg]^{2} \,. \label{eq:eigenvalues_supplementary}
\end{equation}
and their maximal number~$n_{\max }$ is determined by ${2 n_{\text{max}} \leq \sqrt{1 + 4 \mathcal{U} \tilde{\xi}_{\text{s}}^{-2} \kappa_{\text{w}}^{-2}} - 1}$.

We expand the correction~$\delta \Delta$ to the zero-order solution~$\Delta_{0}$ in terms of the normalized eigenfunctions~${\Delta_n \equiv \psi_{n}}$ of the operator ${\hat{\mathcal{L}} = - \tilde{\xi}_{\text{s}}^2 \partial_{xx}^2 - \mathcal{U} \cosh^{-2} (\kappa_{\text{w}} x)}$. These functions obey the equation
\begin{equation}
\hat{\mathcal{L}} \psi_{n} = \mathcal{E}_{n} \psi _{n} \,.  \label{eq:operator_equation_for_psi}
\end{equation}

Solutions of Eq.~\eqref{eq:GP_main} can be written explicitly if the quantity ${\mathcal{E} = \mathcal{E}(\eta, \delta [\mu^{2}])}$ is close to a certain `energy' level~$\mathcal{E}_{n}$, say to~$\mathcal{E}_{N}$, such that ${\mathcal{E} \simeq \mathcal{E}_{N} = \mathcal{E}(\eta_{N}, \delta [\mu_{N}^{2}])}$ (if considering the model for cuprates or iron-based pnictides, the `temperature'~$\eta$ or doping~$\delta [\mu^{2}]$ should be chosen properly). We write Eq.~\eqref{eq:GP_main} in the form
\begin{equation}
\hat{\mathcal{L}} \Delta = \mathcal{E}_{N} \Delta + R(\Delta) \,, \label{eq:GP_with_restterm}
\end{equation}
with ${R = g \Delta^{3} + (\mathcal{E} - \mathcal{E}_{N}) \Delta}$ and represent~$\Delta$ as ${\Delta(x) = c_{N} \psi_{N}(x) + \delta \Delta_{N}(x)}$, where ${\delta \Delta_{N}(x) = \sum_{n}^{\prime} c_{N,n} \psi_{n}(x)}$, and the summation runs over all~$n$ except the term ${n = N}$. We substitute this~$\Delta(x)$ into Eq.~\eqref{eq:GP_with_restterm} and multiply this equation first by~$\psi_{N}$ and then by~$\psi_{n}$ with ${n \neq N}$, then integrating the obtained result each time over~$x$. Thus, taking into account the orthogonality of different eigenfunctions, we find the coefficients~$c_{n}$
\begin{align}
c_{N}^{2} &= \frac{\mathcal{E} - \mathcal{E}_{N}}{g \langle\langle \psi_{N}^{4} \rangle\rangle} \,, \label{eq:c_N} \\
c_{N,n} &= g c_{N}^{3} \frac{\langle\langle \psi_{N}^{3} \psi_{n} \rangle\rangle}{E_{n} - E_{N}} \qquad \text{with ${n \neq N}$} \,,  \label{eq:c_Nn}
\end{align}
where $\langle\langle f(x) \rangle\rangle = \int_{-\infty}^{\infty} \mathrm{d} x \, f(x)$. Obviously, in Eq.~(\ref{eq:c_Nn}),~$\psi_{n}$ and~$\psi_{N}$ have to have same parity (both even or both odd).


%

\end{document}